\documentclass[letterpaper, 10 pt,conference]{ieeeconf}  % Comment this line out if you need a4paper
\usepackage{amssymb}
\usepackage{amsmath,mathrsfs}
\usepackage{color}
\usepackage{graphicx}
\usepackage{comment}
\usepackage{mathtools}
\usepackage{lineno,hyperref}
\usepackage{cleveref}
\usepackage{bm}
\usepackage{tikz}
\usepackage{algorithm}
\usepackage{algorithmicx}
\usepackage{algpseudocode}
\modulolinenumbers[5]

\usepackage{multirow}
\usepackage{color}
\usepackage{cite}

\providecommand{\keywords}[1]{\textbf{\textit{Index terms---}}}

\IEEEoverridecommandlockouts                              % This command is only needed if 
\title{\LARGE \bf Grey-box Recursive Parameter Identification of a Nonlinear Dynamic Model for Mineral Flotation}

\author{Rodrigo A. Gonz\'alez and Paulina Quintanilla%, other coauthors % <-this % stops a space
\thanks{R. Gonz\'alez is with the Department of Mechanical Engineering, Eindhoven University of Technology, The Netherlands. P. Quintanilla is with the Department of Chemical Engineering, Brunel University London, UK. E-mails: r.a.gonzalez@tue.nl; paulina.quintanilla@brunel.ac.uk. (R.  Gonz\'alez and P. Quintanilla are co-first authors).}}%
\begin{document}

\maketitle

%%%%%%%%%%%%%%%%%%%%%%%%%%%%%%%%%%%%%%%%%%%%%%%%%%%%%%%%%%%%%%%%%%%%%%%%%%%%%%%%
\begin{abstract}
This study presents a grey-box recursive identification technique to estimate key parameters in a mineral flotation process across two scenarios. The method is applied to a nonlinear physics-based dynamic model validated at a laboratory scale, allowing real-time updates of two model parameters, $n$ and $C$, in response to changing conditions.
The proposed approach effectively adapts to process variability and allows for continuous adjustments based on operational fluctuations, resulting in a significantly improved estimation of concentrate grade – one key performance indicator. In Scenario 1, parameters $n$ and $C$ achieved fit metrics of 97.99 and 96.86, respectively, with concentrate grade estimations improving from 75.1 to 98.69 using recursive identification. In Scenario 2, the fit metrics for $n$ and $C$ were 96.27 and 95.48, respectively, with the concentrate grade estimations increasing from 96.27 to 99.45 with recursive identification. The results demonstrate the effectiveness of the proposed grey-box recursive identification method in accurately estimating parameters and predicting concentrate grade in a mineral flotation process.
\end{abstract}


\section{Introduction}
Control theory offers an extensive framework for the design of optimal controllers for linear and nonlinear systems. Techniques such as model predictive control (MPC) use a dynamic process model to predict future plant responses and optimize control signals accordingly. In practical applications, certain system parameter values are unknown, and control laws are often determined based solely on available model information. Since model inaccuracies can result in deviations from the optimal solution, parameter estimation in complex dynamical models becomes crucial across various engineering domains \cite{Runge2020}, including mineral processing. 

Mineral flotation process is crucial in extracting valuable minerals from impurities. Modeling mineral flotation presents additional complexities due to its multiphase nature, in which solid (mineral), liquid (water) and gas (air) components are involved. A recent dynamic model incorporating froth physics was developed and validated at laboratory scale \cite{quintanilla2021dynamic, quintanilla2021dynamic2}. This new model was implemented in an economic MPC strategy \cite{Quintanilla2023EconomicModels}, but assumes constant model parameters despite their likely changes under disturbances or varying operating conditions. Accurate identification of unknown parameters in complex dynamical models is essential for effective control strategies and optimizing key performance indicators such as mineral recovery and concentrate grade.

A data-driven approach to deal with the time-varying nature of the dynamic model consists of continuously updating the model parameters as more data is collected. In this regard, the field of recursive identification \cite{ljung1983theory} deals with the problem of designing computationally efficient parameter estimation methods that can track the evolution of a system in real time. These methods have been successfully deployed to estimate a diverse range of physical processes, including electromechanical \cite{pan2019grey}, hydraulic \cite{wigren1993recursive}, and rainfall forecasting models \cite{young2010real}. %Traditional recursive identification algorithms such as recursive least squares or Kalman-based approaches \cite[Chap. 9]{soderstrom1989system} consider discrete-time black-box models, and the model structure is determined empirically or with model order selection methods such as Akaike's information criterion (AIC, \cite{akaike1974new}) and the Bayesian information criterion (BIC, \cite{schwarz1978estimating}). 
%Contrary to these conditions, 
Mineral flotation is typically described by continuous-time nonlinear dynamics derived through first principles modeling. Since the parameters to be estimated using data represent phenomena that are deeply intertwined with the physically-revelant variables, a grey-box modeling philosophy must be adopted \cite{bohlin2006practical}.

The main contribution of this work is the application of a recursive algorithm for the online estimation of two key parameters in a nonlinear dynamic mineral flotation model \cite{quintanilla2021dynamic}. This approach allows adaptation to changing process conditions without manual re-calibration, providing flexibility in responding to varying operational conditions and holding potential for future research in improving froth flotation predictive control. 

In the following sections, we will cover the description of the froth flotation process (Section \ref{sec:processdescription}), the outline of the model equations (Section \ref{sec:modeloutline}), the derivation and implementation aspects of the proposed recursive identification method (Section \ref{sec:grey-box}), the presentation of the results of this methodology in the froth flotation model (Section \ref{sec:results}), and the conclusion of this work (Section \ref{sec:conclusions}).

\section{Process Description}
\label{sec:processdescription}
\begin{figure}[h!]
    \centering
    \hspace{-35pt}
\includegraphics[width=0.55\textwidth] %[scale=.5]%
{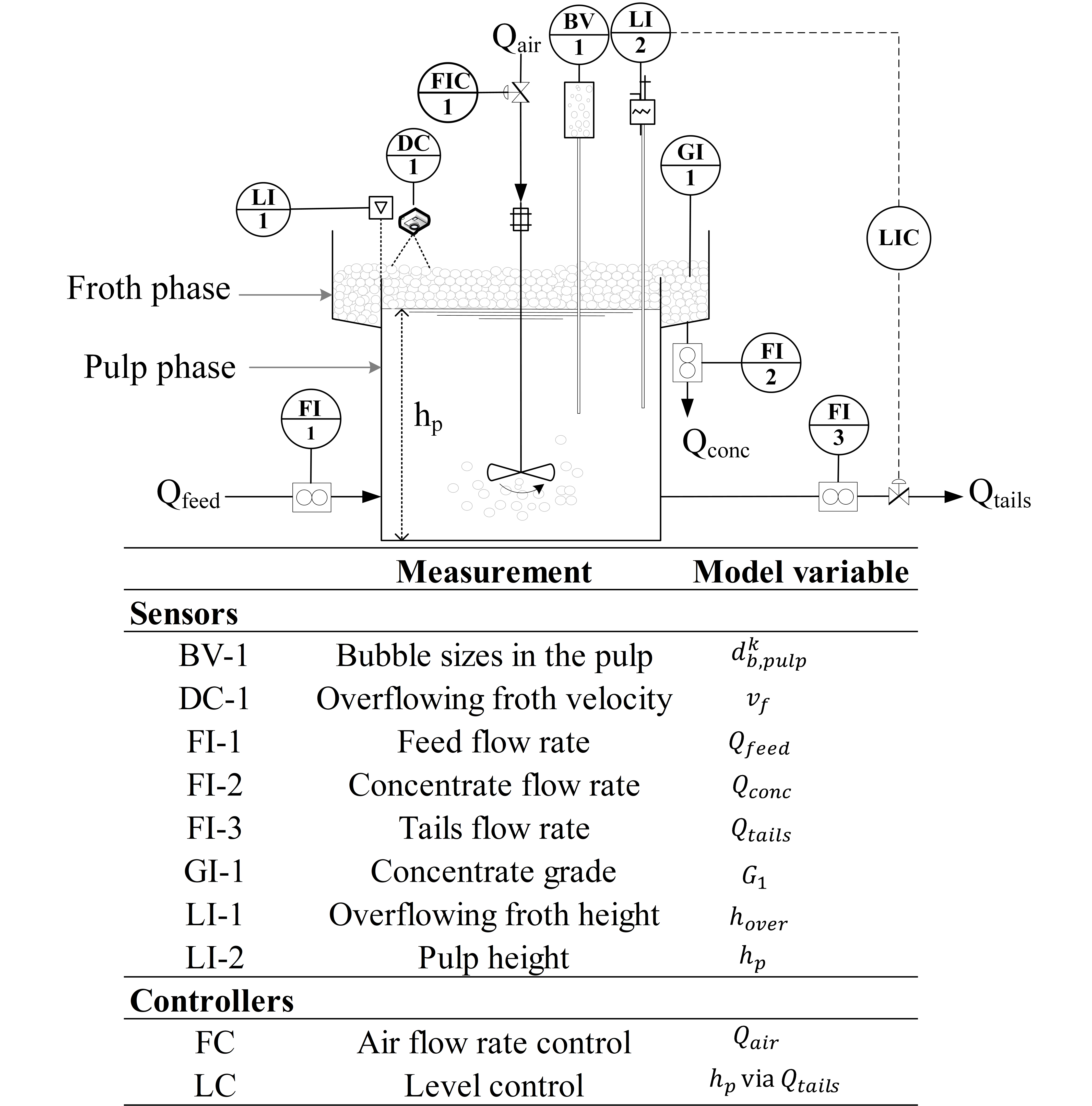}
    \caption{Simplified P\&ID of a froth flotation cell (adapted from \cite{Quintanilla2023EconomicModels}) and the froth flotation phases (froth, interface and pulp). The inlet flow rates, feed ($Q_{feed}$) and air ($Q_{air}$), and outlet flow rates, concentrate ($Q_{conc}$) and tails ($Q_{tails}$) are also indicated. The controlled variable is pulp height ($h_p$), which is controlled by manipulating the tails flow rate.}
    \vspace{-0.5cm}
    \label{fig: froth-flotation}
\end{figure}

Mineral flotation is a process that involves mixing chemicals and air into stirred cells to modify the surface properties of mineral particles, making them water-repellent and causing them to attach to air bubbles. The resulting mixture generates a froth at the cell's surface that overflows as a mineral-rich concentrate. The unattached particles remain in the tank in the pulp phase, and leave from the bottom of the tank as tails. Figure \ref{fig: froth-flotation} illustrates a mineral flotation cell and its standard instrumentation.
In terms of process control, the air flow rate ($Q_{air}$) and the tails flow rate ($Q_{tails}$) are manipulated variables. The tails flow rate is used to regulate the height of the pulp ($h_p$), which is monitored by a level sensor (LI-2). The feed flow rate ($Q_{feed}$) is considered a measurable disturbance coming from upstream processes. The sensor BV-1 is a Bubble Viewer that uses image analysis to calculate the bubble size distribution within the pulp \cite{Mesa2022}.

\section{Model Outline}
\label{sec:modeloutline}
In this study, we consider a novel dynamic model that incorporates the froth physics \cite{quintanilla2021dynamic,quintanilla2021dynamic2}, a key driver of flotation performance. The application of this dynamic model in an economic MPC strategy was discussed in \cite{Quintanilla2023EconomicModels}. However, this approach assumed constant model parameters, which may not accurately capture the changes caused by disturbances and varying operating conditions. To address these modeling limitations, a grey-box recursive identification method is proposed in this paper. 

The model is a nonlinear Differential and Algebraic Equation (DAE) system with $2+I+K$ ordinary differential equations, where $I$ is the number of mineralogical classes and $K$ is the number of bubble size classes. In this study, we assumed $K=5$ and $I=2$, corresponding to chalcopyrite (valuable mineral, $i = 1$) and quartz (waste rock, $i = 2$). %The model has 5 tuning parameters ($n,C,a, b$ and $c$). 
Key elements of the dynamic model are provided in this section. %For a detailed understanding of the model development and its physical interpretation, readers are directed to \cite{quintanilla2021dynamic, quintanilla2021dynamic2}.

The mass conservation for the mineralogical class $i$ is described by \eqref{eq: mass-balance}:
\begin{equation}
\label{eq: mass-balance}
    \frac{\textnormal{d}m_{i}}{\textnormal{d}t}=m_{i, feed}-m_{i, tails}-m_{i, TF}-m_{i, ent}.
\end{equation}
The rate of change of mass for the mineralogical class \textit{i}, $\textnormal{d}m_i/\textnormal{d}t$, is determined by considering the %input 
particles entering into the flotation cell from the feed ($m_{i, feed}$), as well as their distribution in the tails ($m_{i, \text{tails}}$) and concentrate ($m_{i, TF}$ and $m_{i, {ent}}$). Specifically, $m_{i,TF}$ represents the mass flow of particles belonging to mineralogical class \textit{i} that are present in the concentrate due to true flotation, while $m_{ i, ent}$ denotes those particles entrained in the concentrate. Substituting each term on the RHS of \eqref{eq: mass-balance} with $m_{i, feed} = C_{i,f}Q_{feed}$, $m_{i, tails} = C_{i, tails}Q_{tails}$, $m_{i, TF} = V_{cell} P_i S_b R_{f,i} C_{tails,i}$, and $m_{ent,i} = Q_{conc}R_{ent, i} C_{tails,i}$ from Eqs. (10), (11), (51), and (56) in \cite{quintanilla2021dynamic}, respectively, we obtain:

\begin{align}
\label{eq: mass-balance-2}
    &\frac{\textnormal{d}m_{i}}{\textnormal{d}t}\hspace{-0.04cm}= \hspace{-0.04cm}C_{i,f }Q_{feed} -\hspace{-0.07cm}\frac{m_i}{V_{pulp}} Q_{tails} - \hspace{-0.07cm}V_{cell} P_i S_b R_{f,i} C_{tails, i}  \notag \\
    &-Q_{conc} R_{ent, i} C_{tails, i},
\end{align}

\noindent where the inputs are the feed grade $C_{i, feed}$, the feed flow rate $Q_{feed}$, and the concentrate flow rate $Q_{conc}$.
The flotation cell volume $V_{cell}$ and the floatability factor $P_i$ of each mineralogical class are assumed fixed and known. The tails flow rate $Q_{tails}$ is assumed to be a manipulated variable via a Proportional-Integral (PI) controller \cite{Quintanilla2023EconomicModels} to tackle an error $e = h_p - h_p^{SP}$, where $h_p$ is pulp height (see \eqref{eq: hp}) and $h_p^{SP}$ is its setpoint. The PI controller is formulated as:
\begin{equation}
\label{eq: Qtails}
\frac{\textnormal{d}Q_{tails}}{\textnormal{d}t}= k_p \frac{\textnormal{d}h_p}{\textnormal{d}t}-\frac{k_{p}}{\tau_{i}} e,
\end{equation}
\noindent where $k_p$ and $\tau_i$ are the proportional and integral parameters. The dynamic evolution of pulp height is given by~\cite{Quintanilla2023EconomicModels}:
\begin{equation}
\label{eq: hp}
\frac{\textnormal{d}h_p}{\textnormal{d}t}=v_g^*- v_{gas, out}^{total},
\end{equation}

%\begin{equation}
%\label{eq: hp}
%\frac{\textnormal{d}h_p}{\textnormal{d}t}=\frac{Q_{feed}-Q_{tails} - Q_{conc} + Q_{air}}{A_{cell}}- v_{gas, out}^{total},
%\end{equation}
\noindent where  $v_{gas, out}^{total}$ is the gas velocity out of the pulp phase (Eq. (33) in \cite{quintanilla2021dynamic}), and $v_g^*$ is the interfacial gas velocity, defined as:

\begin{equation}
\label{eq: vg*}
    v_g^* = \frac{Q_{feed}-Q_{tails}-Q_{conc}+Q_{air}}{A_{cell}},
\end{equation}
\noindent where the input signal $Q_{air}$ is the air flow rate, and $A_{cell}$ is the cross-sectional area of the flotation cell, which is constant and assumed to be known. Considering that $V_{pulp}= h_0 A_{cell} + V_{gas}$, $V_{gas} = \sum_{k=1}^K \varepsilon_0^k/(1-\varepsilon_0^k)h_0A_{cell}$, and  $S_b = 6v_g^*/d_{b_{int}}$ \cite{quintanilla2021dynamic}, we obtain: %(Eqs. (14), (16), and (53) in \cite{quintanilla2021dynamic}, respectively), we obtain:

\begin{align}
\label{eq: mass-balance-3}
&\frac{\textnormal{d}m_{i}}{\textnormal{d}t}\hspace{-0.03cm}= \hspace{-0.03cm}C_{i,f }Q_{feed} -\Bigg(\frac{m_i}{h_0 A_{cell} \big(1 + \sum^K_{k=1}\frac{\varepsilon_0^k}{1-\varepsilon_0^k}\big)}\Bigg) \notag \\ 
&\hspace{0.3cm} \times \left ( Q_{tails} + \hspace{-0.03cm}V_{cell} P_i \frac{6v_g^*}{d_{b_{int}}} R_{f,i} +\hspace{-0.03cm} Q_{conc}R_{ent, i} \right ),
\end{align}
where $d_{b_{int}}$  is the interfacial bubble size (Eq. (38) in \cite{quintanilla2021dynamic}), and it is described by means of pulp bubble size (input) and $v_{gas, out}^{total}$ (Eq. (32) in \cite{quintanilla2021dynamic}). The term $h_0$ refers to the gas-free pulp level and it is described by:
\begin{equation}
    h_0 = \bigg({1-\sum_{k = 1}^K \varepsilon_0^k}\bigg) h_p.
\end{equation}
The term $\varepsilon_0^k$ is the gas holdup in the pulp phase for the bubble size class $k$. If we denote the proportion of the bubble size class $k$ as $\Psi^k$, and define $\varepsilon_0^{total} = \sum_{k=1}^{K} \varepsilon_0^k$, the dynamics associated with $\varepsilon_0^k$ obey
\begin{align}
\label{eq: gas-holdup}
&\frac{\textnormal{d}\varepsilon_0^{k}}{\textnormal{d}t}=\frac{1+\varepsilon_0^{total}}{A_{cell} h_{p}} \Biggl( Q_{air} \Psi^{k}-A_{cell} v_{gas, out}^{k} \frac{\varepsilon_0^{k}}{1+\varepsilon_0^{total}} \notag \\
&-(Q_{feed}-Q_{tails} - Q_{conc})\varepsilon_0^{k} \Biggr)  \quad \text{for } k = 1, \ldots, K.
\end{align}

A critical aspect of improving the flotation process and mineral recovery rates is understanding the froth's behavior, such as its stability and the interactions between bubbles and particles. A key metric of froth stability is air recovery, which correlates closely with the efficiency of the flotation process \cite{Hadler2010}. Air recovery is calculated in real-time using a digital camera to capture the speed of the froth overflowing the cell top ($v_f$), a level sensor to measure the froth's height at the cell's edge ($h_{over}$), and the rate of air flow ($Q_{air}$). The formula to determine the air recovery ($\alpha$) is given by:

\begin{equation}
    \alpha = \frac{v_f h_{over} l_{lip}}{Q_{air}},
\end{equation}

\noindent where $l_{lip}$ represents the known perimeter of the cell.

The term $R_{f, i}$ in \eqref{eq: mass-balance-3} is the froth recovery, defined as the fraction of mineral particles entering the froth phase attached to the bubbles that overflow as concentrate, rather than dropping back to the pulp phase. A approximation of froth recovery was defined by \cite{Neethling2008} as:

\begin{equation}
\label{eq: rf}
R_{f,i}\hspace{-0.07cm}=\hspace{-0.07cm} \begin{cases}
        \left(\frac{\alpha^{*}\left(1-\alpha^{*}\right)   v_{g}^{*}}{v_{{set}, i}}\right)^{\hspace{-0.04cm}0.25}\left(\frac{d_{b, {int}}}{d_{{b, froth_{out}}}}\right)^{\hspace{-0.04cm}0.5} & \hspace{-0.3cm}\text{if } \alpha\hspace{-0.03cm}<\hspace{-0.03cm}0.5 \\
        \left(\frac{v_{g}^{*}}{4 v_{{set},   i}}\right)^{0.25}\left(\frac{d_{b,{ int }}}{d_{{b, froth_{out}}}}\right)^{0.5} & \hspace{-0.3cm}\text{if } \alpha\hspace{-0.03cm}\geq \hspace{-0.03cm}0.5
    \end{cases} 
\end{equation}
\noindent where $v_{set,i}$ is the settling velocity (Eq. (55) in \cite{quintanilla2021dynamic}), and $\alpha^* = 1 - {v_b}/{v_g^*}$, with $v_b$ being the bursting rate from \cite{Neethling2018}.

The term $d_{{b, froth}_{out}}$ is the bubble size on the top of the froth phase \cite{Neethling2008}, that is represented by:

\begin{equation}
\label{eq: db-froth}
    d_{{b, froth}_{out}} = (n C \tau_f + d_{b, int}^n)^{1/n}, 
\end{equation}
\noindent where $n$ and $C$ are parameters of the model, and $\tau_f$ is the froth residence time that is quantified as $\tau_f = (h_T-h_p)/v_g^*$, where $h_T$ is the total height of the flotation cell (constant, assumed to be known) and $v_g^*$ is described by \eqref{eq: vg*}. 

%The term $\alpha^*$ is the definition of air recovery using the actual bursting rate ($v_b$) from  \cite{Neethling2018} and the interfacial gas velocity ($v_g^*$) from \cite{quintanilla2021dynamic}. 

\begin{comment}
The quadratic relationship between $v_b$ and $Q_{air}$ is derived from industrial data analysis (more details in \cite{Neethling2018}). Taking \eqref{eq: vg*} and defining

\begin{equation}
    \label{eq: vb}
    v_b = a + b \left(\frac{Q_{air}}{A_{cell}}\right)+c \left(\frac{Q_{air}}{A_{cell}}\right)^2,
\end{equation}
   
\end{comment}

%\noindent where $a$, $b$ and $c$ are tuning parameters. 

The term $R_{ent, i}$ in \eqref{eq: mass-balance-3} is the entrainment factor, corresponding to the proportion between the amount of waste rock entrained in the froth phase and the water recovery. A simplified model for $R_{ent, i}$ was defined by  \cite{Neethling2009} as:
\begin{align}
\label{eq: ent}
    R_{ent, i} &= \begin{cases}
        \exp \left(\frac{-v_{set, i}^{1.5}   (h_T - h_p)}{D_{{axial}} \sqrt{v_{g}^{*}\left(1-\alpha^{*}\right)}}\right) & \text{if } \alpha<0.5 \\
        \exp \left(\frac{-2 v_{set, i}^{1.5}   (h_T - h_p)}{D_{ {axial}} \sqrt{v_{g}^{*}}}\right) & \text{if } \alpha\geq 0.5
    \end{cases} 
\end{align}
\noindent where  $D_{axial}$ is a function of air flow rate ($Q_{air}$) as Eq. (58) in \cite{quintanilla2021dynamic}. 
The concentrate flow rate, $Q_{conc}$, is given by \cite{Neethling2003}:

\begin{align}
\label{eq: Qconc}
Q_{conc} &= \begin{cases}
        \frac {6.815 A_{cell}   {v_g^*}^{2}(1-\alpha^*) \alpha^*}{k_{1} d_{b, froth_{out}}^2} & \text{if } \alpha<0.5 \\
\frac{6.815 A_{cell} {v_g^*}^{2}}{4 k_{1} d_{b, froth_{out}}^2} & \text{if } \alpha\geq 0.5
\end{cases}
\end{align}

\noindent where $k_1$  (Eq. (45) in\cite{quintanilla2021dynamic}) is a physical parameter given by means of operating conditions such as pulp density and pulp viscosity (Eqs. (34) and (36) in \cite{quintanilla2021dynamic}, respectively). %Recent developments in instrumentation have demonstrated their practicality and effectiveness in delivering real-time assessments of concentrate flow rate \cite{Finfer2018}. Nonetheless, it is acknowledged that online measurement of this variable at an industrial level is not widespread yet.

The flotation process relies on two key performance indicators: concentrate grade and recovery. Concentrate grade reflects the quality or purity of the produced concentrate, while recovery indicates the percentage of valuable minerals successfully extracted from the ore. Online measurement is used to determine concentrate grades by assessing mineral concentration in the stream, whereas recovery measurement is complex and often involves indirect, infrequent estimations. The phenomenological model for estimating concentrate grade of the valuable mineral ($G_1$) is:

\begin{equation}
\label{grade}
    G_1 = \frac{m_{1, TF} + m_{1, ent}}{\sum^2_{i=1} m_{i, TF} + m_{i, ent}},
\end{equation}

\noindent where $i = 1$ corresponds to the valuable mineral (chalcopyrite) and $i=2$ is the gangue (quartz). Each term can be replaced as in Eq. \eqref{eq: mass-balance-2}, yielding:

\begin{equation}
\label{eq:grade}
    G_1 = \frac{C_{tails, 1}(V_{cell} P_1 S_b R_{f,1}  + Q_{conc} R_{ent, 1})}{\sum^2_{i=1} C_{tails, i}(V_{cell} P_i S_b R_{f,i} + Q_{conc} R_{ent, i})}.
\end{equation}

Note that $R_{f,i}$ in \eqref{eq: rf} and $Q_{conc}$ in \eqref{eq: Qconc} involve the model parameters $n$ and $C$ from $d_{b, froth_{out}}$ in \eqref{eq: db-froth}. %These parameters play a critical role in defining the mass conservation calculations in \eqref{eq: mass-balance-3} which are used to determine KPIs like recovery and concentrate grade (Eqs. (16) and (18) in \cite{Quintanilla2023EconomicModels}, respectively). 
Accurate estimation of these parameters from data is essential, and they should be updated when there are changes in operating conditions. To tackle this challenge, we propose employing a grey-box recursive identification of the model, which allows for better adaptation to changing system conditions.

\begin{comment}
    Rearranging the terms in \eqref{eq: rf} and \eqref{eq: ent} using \eqref{eq: vg*} and \eqref{eq: alpha*}, we obtain new equations in terms of the tuning parameters $a, b,$ and $c$:

\begin{equation}
\label{eq: rf}
R_{f, i}=\left\{\begin{array}{c}\frac{\left(1-\frac{A_{cell}(a + b \frac{Q_{air}}{A_{cell}}+c \left (\frac{Q_{air}}{A_{cell}} \right )^2}{Q_{feed}-Q_{tails}-Q_{conc}+Q_{air}} \right)^{f/2} \left (a + b \frac{Q_{air}}{A_{cell}}+c \left (\frac{Q_{air}}{A_{cell}} \right )^2 \right)^{f/2}}{v_{\text {set }, i}^{f/2}} \left(\frac{d_{b, \text { int }}}{d_{\text {bfroth }}}\right)^{f} \text { if } \alpha<0.5 \\ 

\left(\frac{Q_{feed}-Q_{tails}-Q_{conc}+Q_{air}}{4 A_{cell}v_{\text {set}, i}}\right)^{f/2}\left(\frac{d_{b, \text { int }}}{d_{\text {bfroth }}}\right)^{f} \text { if } \alpha \geq 0.5\end{array}\right.
\end{equation}

\end{comment}

%Conservacion de masa (implica escribir $R_{f,i}, R_{ent, i}$... tal vez conviene tirar alguna de estas ecuaciones directamente al apendice?)
%Derivadas de $\epsilon_0^k$
%Ecuacion para $h_p$
% Ecuacion para $\alpha$ (y/o $\alpha*$)
%Ecuacion para $Q_{conc}\implies$ deducir que se puede obtener $Q_{conc}$ en terminos de las otras variables

\section{Grey-box Recursive Identification}
\label{sec:grey-box}

In this section, we report the experimental data and the identification method that will be used to estimate in real-time the parameters of interest $n$ and $C$ in \eqref{eq: db-froth}. As seen from the froth flotation model described in Section \ref{sec:modeloutline}, the concentrate grade $G_1$ contains the dependencies on the parameters to be estimated, and is thus considered as the measurable output signal of our model for identification.

\begin{comment}
    More precisely, the model $\mathcal{G}_{\bm{\theta}}$ is described by
\begin{equation}
\label{model}
\mathcal{G}_{\bm{\theta}}\colon \begin{cases} \displaystyle\frac{\textnormal{d}\mathbf{x}(t)}{\textnormal{d}t} \hspace{-0.35cm}&= f_{\bm{\theta}}(\mathbf{x}(t),\mathbf{u}(t)) \\
	\hspace{0.25cm}\mathbf{y}(t)\hspace{-0.5cm} &= g_{\bm{\theta}}(\mathbf{y}(t),\mathbf{x}(t),\mathbf{u}(t)),
\end{cases}
\end{equation}
where $\mathbf{x}(t)\in \mathbb{R}^n$ is the state vector, $\mathbf{u}(t)\in \mathbb{R}^m$ is the input vector, and $\mathbf{y}(t)\in \mathbb{R}^p$ is the output vector. For simplicity, we assume that the input vector also contains the measured disturbances of the system, and that the disturbances are measured without error.

Contrary to most nonlinear state-space representations used in system identification \cite{ljung1998system}, the output vector of the grey-box model we intend to identify in \eqref{model} cannot be expressed directly in terms of the state and input vectors only. Although such direct description will exist for most practical cases thanks to the implicit function theorem, we are interested in working with $g_{\bm{\theta}}$ instead, since this function conveys the physical interpretation that stems from the modelling of the process. 
\end{comment}
We assume that a noisy measurement $G_{m1}$ of the true concentrate grade $G_1$ is obtained at the time instants $t=t_1,t_2\dots,t_N$, where the sampling instants $t_k (k=1,\dots,N)$ are assumed to be evenly spaced in time. That is,
\begin{equation}
G_{m1}(t_k) = G_1(t_k) + e(t_k), \quad k = 1,\dots,N,
\end{equation}
where $\{e(t_k)\}_{k=1}^N$ is a white noise stochastic process.

Next, we derive a recursive estimator for $\bm{\theta}=[n,C]^\top$, using the available data $Q_{air}$, $Q_{feed}$, $h_p^{SP}$, and the output $G_1$ at the time instants $t_1,t_2,\dots,t_N$, with increasing $N$ as more data is retrieved. The estimator will be denoted as $\hat{\bm{\theta}}_N$ (i.e., the estimate of $\bm{\theta}$ using the data up to $t=t_N$).

\subsection{Recursive Prediction Error Method}

A natural approach towards obtaining $\hat{\bm{\theta}}_N$ consists in studying the minimization of the prediction error cost
\begin{equation}
V_N(\bm{\theta}) =  \frac{1}{2}\sum_{k=1}^N\lambda^{N-k}\eta^2(t_k,\bm{\theta}),
\end{equation}
where $\eta(t_k,\bm{\theta})$ denotes the output residual $G_{m1}(t_k)-G_1(t_k,\bm{\theta})$, and $\lambda\in(0,1]$ is a forgetting factor. In addition, we have made explicit the dependence of $\bm{\theta}$ in the concentrate grade $G_1(t_k,\bm{\theta})$, which is computed using \eqref{eq:grade}. Note that this computation requires solving the DAE, i.e., the state ordinary differential equations in \eqref{eq: Qtails}, \eqref{eq: hp}, \eqref{eq: mass-balance-3}, and \eqref{eq: gas-holdup}, and the algebraic relations also described in Section \ref{sec:modeloutline}.

Recursive estimators typically assume that an initial estimate $\hat{\bm{\theta}}_{N-1}$ is fixed. This estimate can be obtained offline as an initial guess or estimate from another dataset, or it can be obtained using the first $N-1$ data points. The recursive prediction error method is based on the 2nd order Taylor expansion of $V_N(\bm{\theta})$, which is given by
\begin{align}
    &V_N(\bm{\theta})= V_N(\hat{\bm{\theta}}_{N-1}) + \frac{\partial V_N(\hat{\bm{\theta}}_{N-1})}{\partial \bm{\theta}^\top} (\bm{\theta}-\hat{\bm{\theta}}_{N-1}) \notag \\
    &+ \hspace{-0.07cm}\frac{1}{2}\hspace{-0.02cm}(\bm{\theta}\hspace{-0.05cm}-\hspace{-0.05cm}\hat{\bm{\theta}}_{\hspace{-0.02cm}N\hspace{-0.02cm}-\hspace{-0.02cm}1})^{\hspace{-0.03cm}\top} \frac{\partial^2 V_{\hspace{-0.02cm}N}\hspace{-0.02cm}(\hat{\bm{\theta}}_{\hspace{-0.02cm}N\hspace{-0.02cm}-\hspace{-0.02cm}1})}{\partial \bm{\theta}\partial \bm{\theta}^\top} (\bm{\theta}\hspace{-0.05cm}-\hspace{-0.05cm}\hat{\bm{\theta}}_{N\hspace{-0.02cm}-\hspace{-0.02cm}1}) \hspace{-0.03cm}+\hspace{-0.03cm} o(\|\bm{\theta}\hspace{-0.05cm}-\hspace{-0.05cm}\hat{\bm{\theta}}_{N\hspace{-0.02cm}-\hspace{-0.02cm}1}\|^2),
\end{align}
where $o(\|x\|^2)$ represents a function such that $o(\|x\|^2)/\|x\|^2\to 0$ as $x\to 0$. If we neglect the higher order terms, which is reasonable if $\hat{\bm{\theta}}_{N}$ does not deviate much from $\hat{\bm{\theta}}_{N-1}$, then setting $\partial V_N (\hat{\bm{\theta}}_N)/\partial \bm{\theta}$ to zero gives the Gauss-Newton update
\begin{equation}
\label{gaussnewton}
\hat{\bm{\theta}}_N = \hat{\bm{\theta}}_{N-1} - \left[\frac{\partial^2 V_{\hspace{-0.02cm}N}\hspace{-0.02cm}(\hat{\bm{\theta}}_{\hspace{-0.02cm}N\hspace{-0.02cm}-\hspace{-0.02cm}1})}{\partial \bm{\theta}\partial \bm{\theta}^\top}\right]^{-1} \frac{\partial V_N(\hat{\bm{\theta}}_{N-1})}{\partial \bm{\theta}}.   
\end{equation}
More explicitly, after defining the vector $\bm{\psi}(t_k,\bm{\theta})=\partial y(t_k,\bm{\theta})/\partial \bm{\theta}$, we can compute the gradient of $V_N$ by following standard matrix calculus rules \cite{magnus2019matrix} as
\begin{align}
    \frac{\partial V_N(\hat{\bm{\theta}}_{N-1})}{\partial \bm{\theta}} &= -\sum_{k=1}^N \lambda^{N-k}\bm{\psi}(t_k,\hat{\bm{\theta}}_{N-1}) \eta(t_k,\hat{\bm{\theta}}_{N-1}) \notag \\
    \label{gradientV}
    &=\hspace{-0.03cm} \lambda\frac{\partial V_{\hspace{-0.02cm}N\hspace{-0.02cm}-\hspace{-0.02cm}1}\hspace{-0.02cm}(\hat{\bm{\theta}}_{\hspace{-0.02cm}N\hspace{-0.02cm}-\hspace{-0.02cm}1})}{\partial \bm{\theta}} \hspace{-0.06cm}-\hspace{-0.04cm} \bm{\psi}(t_{\hspace{-0.01cm}N}\hspace{-0.02cm},\hspace{-0.02cm}\hat{\bm{\theta}}_{\hspace{-0.02cm}N\hspace{-0.02cm}-\hspace{-0.02cm}1}) \eta(t_{\hspace{-0.01cm}N}\hspace{-0.02cm},\hspace{-0.02cm}\hat{\bm{\theta}}_{\hspace{-0.02cm}N\hspace{-0.02cm}-\hspace{-0.02cm}1}) \\
    &=-\bm{\psi}(t_N,\hat{\bm{\theta}}_{N-1}) \eta(t_N,\hat{\bm{\theta}}_{N-1}), \notag
\end{align}
where we have assumed that $\hat{\bm{\theta}}_{N-1}$ minimizes the cost $V_{N-1}(\bm{\theta})$, which gives $\partial V_{N-1}(\hat{\bm{\theta}}_{N-1})/\partial \bm{\theta} = \mathbf{0}$. Moreover, using the derivation in \eqref{gradientV}, we have
\begin{align}
    &\hspace{-0.3cm}\frac{\partial^2 V_{N}(\hat{\bm{\theta}}_{N-1})}{\partial \bm{\theta}\partial \bm{\theta}^\top} \notag \\
    \label{antiwoodbury}
    &\hspace{-0.25cm}\approx \lambda \frac{\partial^2 V_{N-1}(\hat{\bm{\theta}}_{N-1})}{\partial \bm{\theta}\partial \bm{\theta}^\top}+\bm{\psi}(t_N,\hat{\bm{\theta}}_{N-1})\bm{\psi}^{\top}(t_N,\hat{\bm{\theta}}_{N-1}),
\end{align}
where we have disregarded the term $[\partial^2 \eta/ \partial \bm{\theta} \partial\bm{\theta}^\top ]\eta$. This term is typically set to zero, since the output residual $\eta$ will be approximately white noise that is independent of the Hessian term close to the true value of $\bm{\theta}$ \cite{ljung1983theory}.

If we define the inverse of the approximate Hessian of $V_N$ as $\mathbf{L}_N$, then \eqref{gaussnewton} reduces to
\begin{equation}
\label{thetaupdate}
    \hat{\bm{\theta}}_N = \hat{\bm{\theta}}_{N-1} + \mathbf{L}_{N} \bm{\psi}_N \eta_N,
\end{equation}
where we have introduced the simplified notation $\bm{\psi}_N:=\bm{\psi}(t_N,\hat{\bm{\theta}}_{N-1})$ and $\eta_N:=\eta(t_N,\hat{\bm{\theta}}_{N-1})$. The matrix $\mathbf{L}_N$ can be computed from $\mathbf{L}_{N-1}$ by applying the Woodbury matrix identity \cite[Sec. 0.7.4]{horn2012} to \eqref{antiwoodbury}:
\begin{align}
\label{lupdate}
\hspace{-0.4cm}\mathbf{L}_N &= \frac{1}{\lambda}\mathbf{L}_{N-1} \notag \\
&\hspace{-0.4cm}- \frac{1}{\lambda}\mathbf{L}_{N-1}\bm{\psi}_N(\lambda+\bm{\psi}_N^\top \mathbf{L}_{N-1}\bm{\psi}_N)^{-1} \bm{\psi}_N^\top \mathbf{L}_{N-1}.
\end{align}

\subsection{Implementation Aspects and Tuning}
\label{subsec:implementation}
The proposed recursive identification procedure in Eqs. \eqref{thetaupdate} and \eqref{lupdate}, involves computing the output residual $\eta_N$ and its gradient $\bm{\psi}_N$. At each time step $t=t_{N}$, we use the state variables of the previous time instant $t_{N-1}$ as initial conditions to solve the DAEs with $\bm{\theta}=\hat{\bm{\theta}}_{N-1}$ as the parameter vector. This yields the predicted concentrate grade $G_1(t_k,{\bm{\theta}}_{N-1})$. Note that it is not required to simulate the model starting at $t=t_1$ at each time step, since all the prior information is encapsulated in the state variables $Q_{tails}$, $h_p$, $m_i$ and $\varepsilon_0^k$. Similarly, each element $j\in \{1,2\}$ of the gradient vector $\bm{\psi}_N$ is computed via second-order numerical differentiation. This procedure requires computing the one step ahead predictors $G_1(t_k,{\bm{\theta}}_{N-1}+\epsilon \mathbf{e}_j)$, with $e_j$ being the $j$th column of the identity matrix of dimension 2, and $\epsilon\ll 1$.

With regards to the tuning parameters of the proposed recursive identification method, we only need to consider the forgetting factor $\lambda$ and the inverse Hessian matrix, $\mathbf{L}_0$. A common rule of thumb is to pick $\lambda$ assuming that the parameters to be estimated remain relatively constant over a time period of $1/(1-\lambda)$ \cite{ljung1983theory}. For our specific application, we have chosen $\lambda$ between 0.99 and 0.995. On the other hand, $\mathbf{L}_0$ is chosen proportional to the empirical covariance of the parameter vector, and $\mathbf{L}_N$ can be retuned depending on the expected variability of each parameter if needed.

\section{Results and Discussions}
\label{sec:results}

We evaluate the effectiveness of the identification method through its impact on the concentrate grade, a measure of the product quality. The performance of the recursive parameter estimation method in improving flotation grade was assessed by comparing its results to both the ground truth and a scenario using constant parameters. 

We consider a case study experiment of a total time duration of $2400$[s], and a sampling period of $1$[s]. The measured concentration grade is contaminated by a Gaussian white noise of standard deviation $10^{-3}$. The first $600$ seconds of the experiment are used for computing the initial estimate of the nominal $n$ and $C$ values for the proposed recursive estimator in an offline fashion, using the prediction error method \cite{ljung1998system}. For $t\in [600,1200]$, the $C$ parameter increases by $20\%$, while $n$ remains constant. For $t\in[1201,1800]$, the $n$ parameter increases by $20\%$, and the last $600$ seconds do not register parameter variations. The nominal values for these parameters are $n=1$ and $C=6.38\cdot 10^{-4}$. The proposed estimation method considers $\lambda=0.995$, and $\mathbf{L}_0$ is tuned as described in Section \ref{subsec:implementation}. 

\begin{figure}[h!]
    \centering
    \includegraphics[scale=0.47]{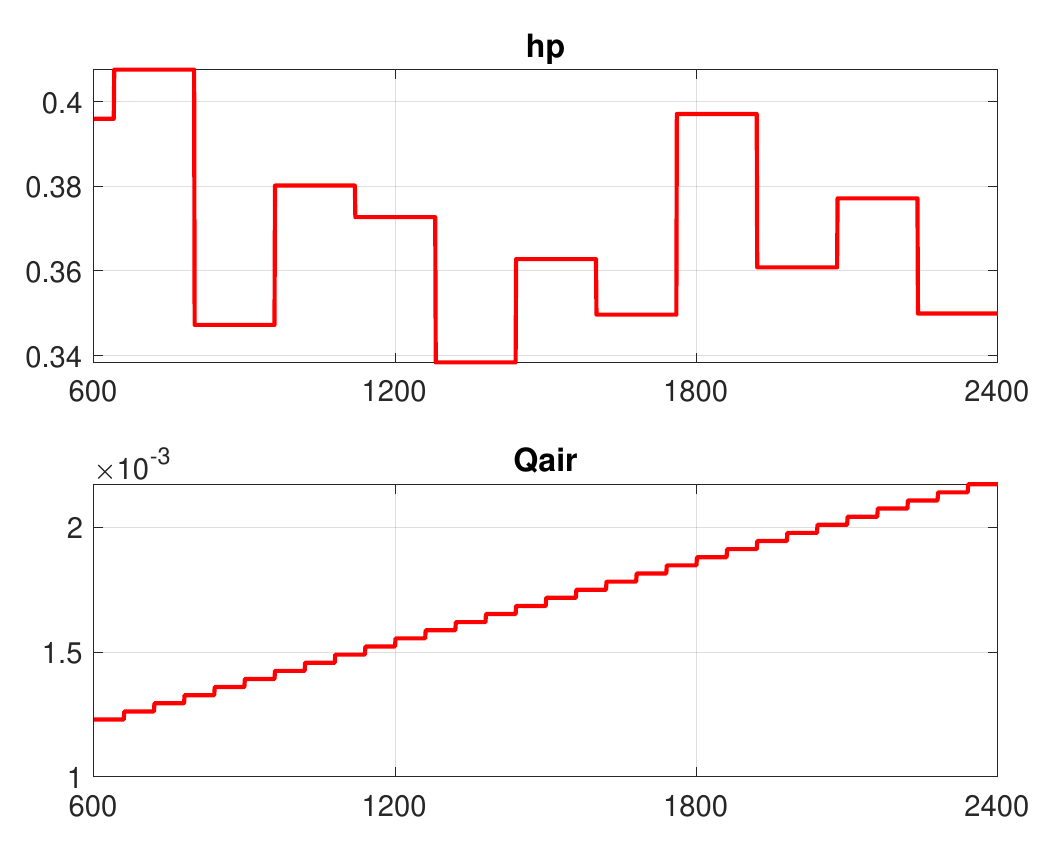}
    \vspace{-0.2cm}
    \caption{Scenario 1: Plots of the inputs $h_p$ and $Q_{air}$ between $600$ and $2400$ seconds.}
    \label{fig:plot-inputs}
    \vspace{-0.5cm}
\end{figure}

\begin{figure}[h!]
    \centering
    \includegraphics[scale=0.47]{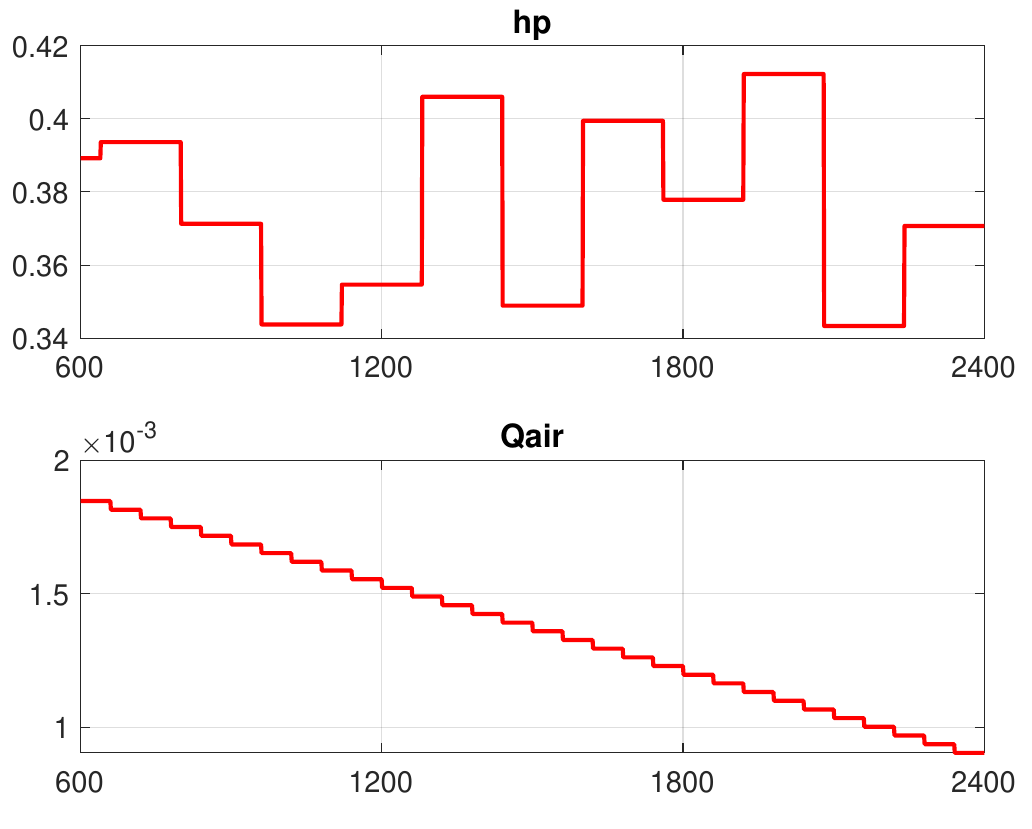}
    \vspace{-0.2cm}
    \caption{Scenario 2: Plots of the inputs $h_p$ and $Q_{air}$ between $600$ and $2400$ seconds.}
    \label{fig:plot-inputs-qairdown}
    \vspace{-0.2cm}
\end{figure}

The control inputs, $Q_{air}$ and $h_p$, are depicted in Figures \ref{fig:plot-inputs} and \ref{fig:plot-inputs-qairdown}. They were designed with amplitude limitations of $0.33$[m] and $0.42$[m] for $h_p$, and $9.05\cdot 10^{-4}$[m\textsuperscript{3}/h] and $2.17 \cdot 10^{-2}$[m\textsuperscript{3}/h] for $Q_{air}$. These limits were selected to align with the operating conditions of the experimental model validation detailed in \cite{quintanilla2021dynamic2}. The change frequency of $h_p$ is set at every 160 seconds, while $Q_{air}$ varies every 60 seconds. Two scenarios were evaluated: (1) $Q_{air}$ increasing; (2) $Q_{air}$ decreasing. In both scenarios, $h_p$ was varied randomly.  


\begin{figure}[h!]
    \centering
    \includegraphics[scale=0.5]{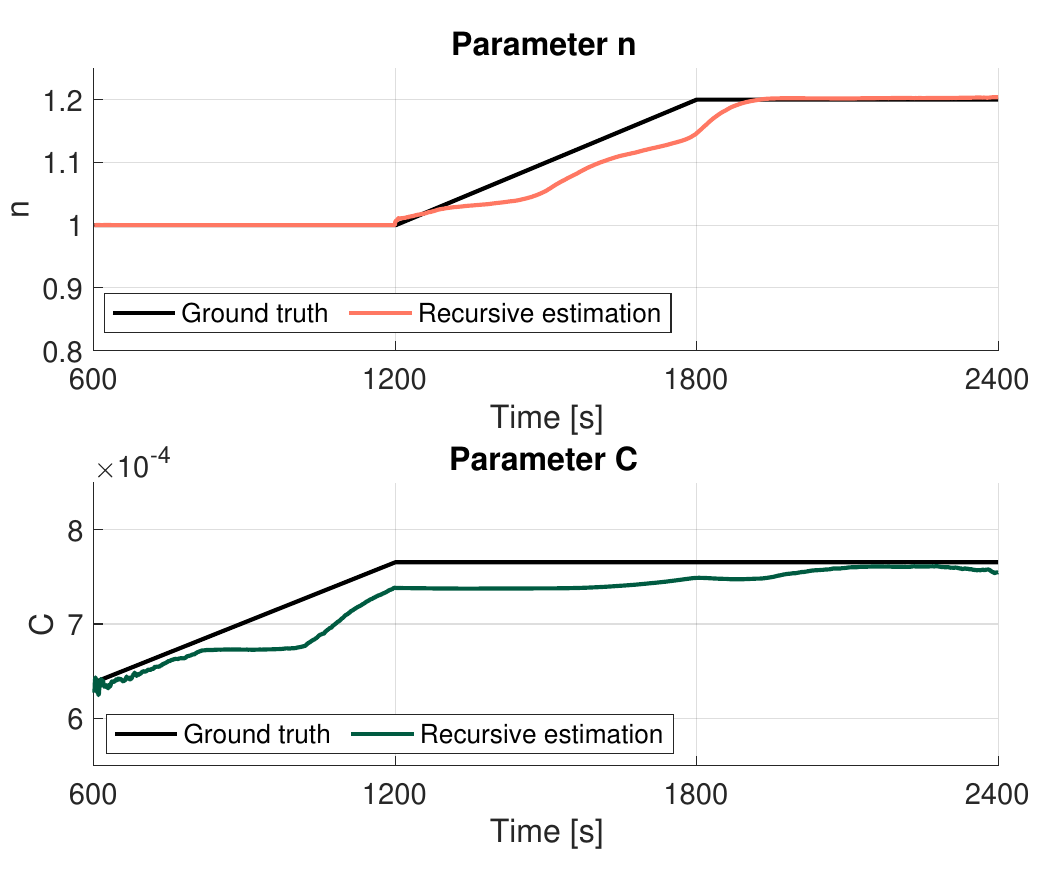}
    \vspace{-1cm}
    \caption{Scenario 1: Recursive estimation of parameters $n$ and $C$ from \eqref{eq: db-froth} over time.}
    \label{fig:plot-nC}
    \vspace{-0.2cm}
\end{figure}

\Cref{fig:plot-nC} illustrates the changes in parameters $n$ and $C$ over time, comparing them to their respective true values. The recursive estimation of the parameter $n$ closely follows the true value throughout the experiment. The estimation is compared qualitatively using a fit metric defined as $100(1-\|\bm{\theta}_i-\hat{\bm{\theta}}_i\|_2/\|\bm{\theta}_i\|_2)$, $i=1,2$, where $\bm{\theta}_1$ and $\bm{\theta}_2$ are vectors containing the respective values of $n$ and $C$ at all time instants after $t=600$ seconds. In this scenario, the fit metric for $n$ is 97.99. The estimation of parameter $C$ similarly exhibits good agreement with the true value, with a fit metric of 96.86. These results reinforce the capability of the proposed recursive method to maintain accuracy in parameter estimation under varying operating conditions.

\begin{figure}[h!]
    \centering
    \includegraphics[scale=0.44]{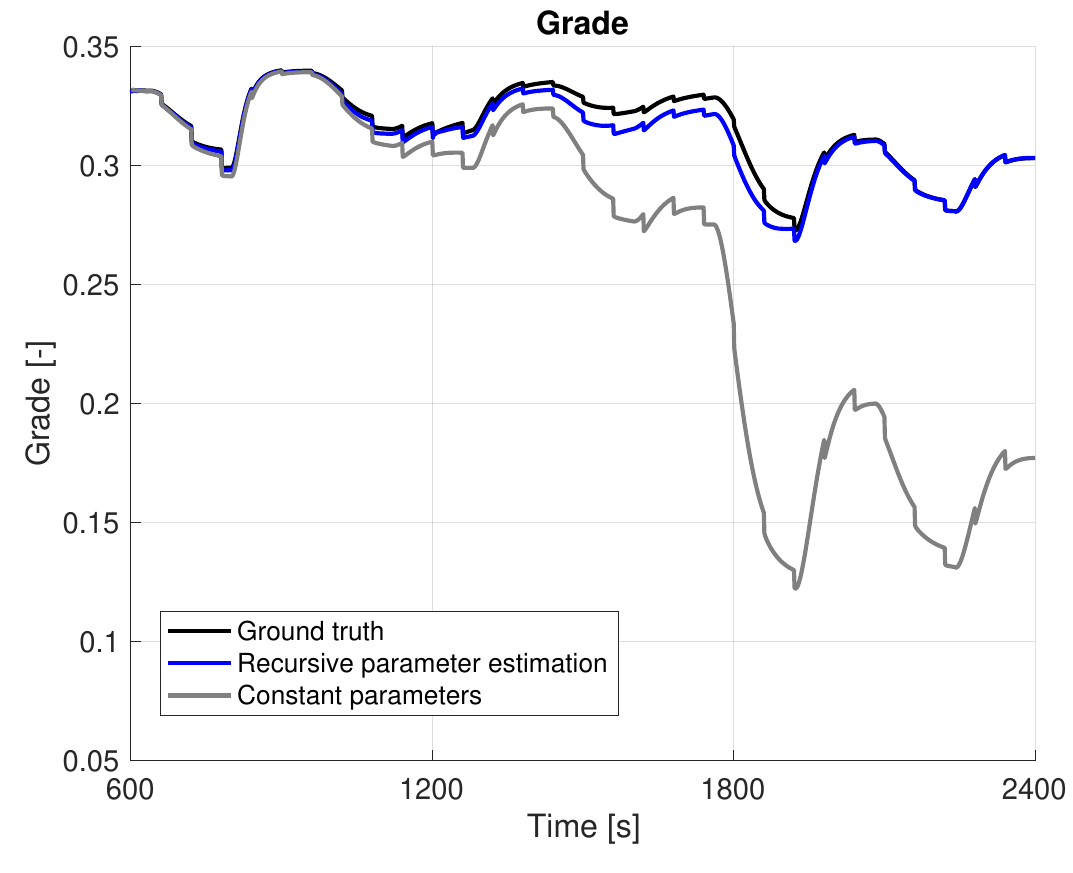}
    \vspace{-0.3cm}
    \caption{Scenario 1: Real-time predicted concentration grade (blue), prediction using nominal parameters (grey), and true concentration grade (black line).}
    \label{fig:plot-grade}
    \vspace{-0.5cm}
\end{figure}

\Cref{fig:plot-grade} shows the evolution of concentrate grade, estimated using the proposed recursive grey-box identification method (blue line) and constant parameters (grey line). The fit metric for concentrate grade is defined similarly to the fit for $n$ and $C$. The grade estimation from the proposed identification method closely matches the true values, with a fit metric of 98.69 for the grey identification method, indicating higher accuracy in capturing mineral concentration fluctuations. In contrast, the constant parameter model deviates noticeably from the ground truth as the experiment progresses. The fit metric using constant parameters is 75.1, significantly lower than that of the online estimation model.

\begin{figure}[h!]
    \centering
    \includegraphics[scale=0.44]{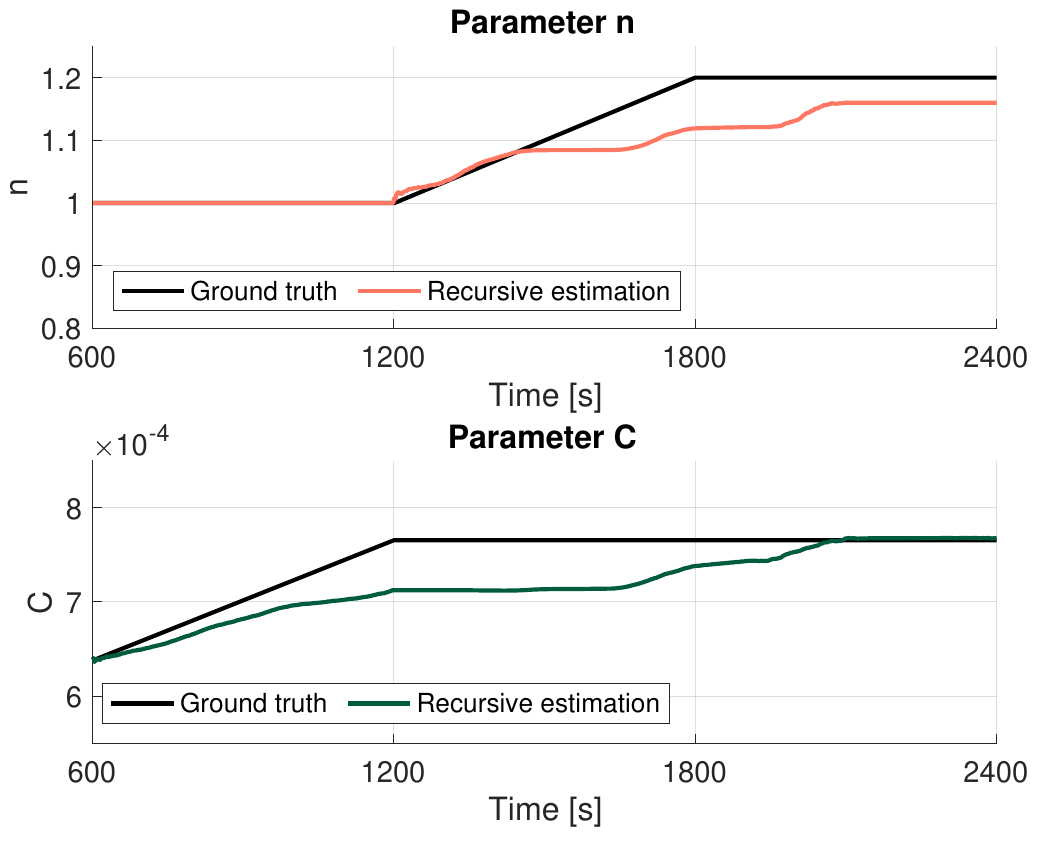}
    \vspace{-0.4cm}
    \caption{Scenario 2: Recursive estimation of parameters $n$ and $C$ from \eqref{eq: db-froth} over time.}
    \label{fig:plot-nC-qairdown}
    \vspace{-0.8cm}
\end{figure}

\begin{figure}[h!]
    \centering
    \includegraphics[scale=0.44]{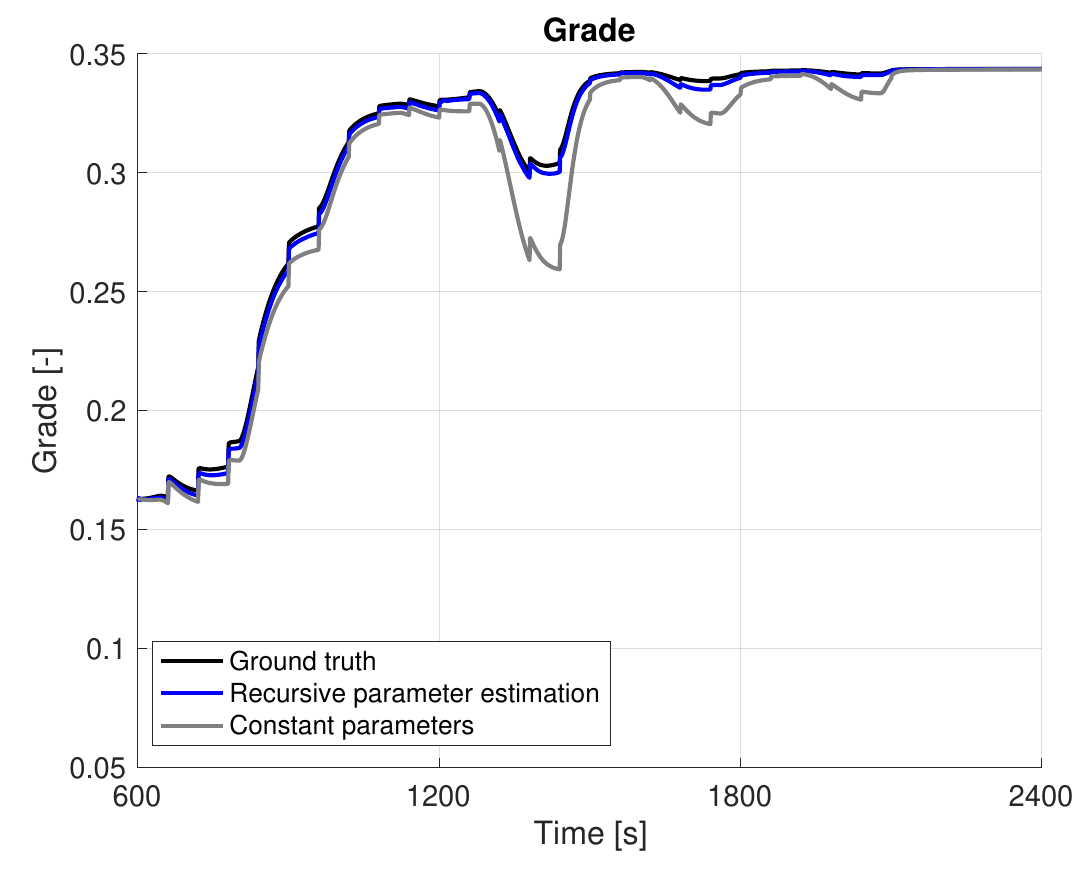}
    \vspace{-0.5cm}
    \caption{Scenario 2: Real-time predicted concentration grade (blue), prediction using nominal parameters (grey), and true concentration grade (black line).}
    \label{fig:plot-grade-qairdown}
    \vspace{-0.5cm}
\end{figure}

Figures \ref{fig:plot-nC-qairdown} and \ref{fig:plot-grade-qairdown} further illustrate the efficacy of the recursive grey-box identification method under varying conditions, with the inputs shown in \Cref{fig:plot-inputs-qairdown}. In \Cref{fig:plot-nC-qairdown}, the recursive estimation of parameters $n$ and $C$ demonstrates fit metrics of 96.27 and 95.48, respectively. Although these errors are slightly higher than those presented in \Cref{fig:plot-nC}, they still underscore a considerable level of accuracy in the parameter estimation process within the different experimental setups.
\Cref{fig:plot-grade-qairdown} compares the accuracy of the recursive method to a constant parameter model (with nominal values in the parameters) in predicting real-time concentration grades. The recursive method shows exceptional precision with a fit metric of 99.45, which is lower than the error observed with the constant parameter model, which stands at 96.27. 
These findings reinforce the earlier observations from Figures \ref{fig:plot-nC} and \ref{fig:plot-grade}, showcasing the recursive method's robust performance across various scenarios. This recursive estimation methodology has consistently low error rates in parameter estimation and concentration prediction, confirming its suitability for complex industrial applications where real-time data and rapid adjustments are critical. 

\section{Conclusions}
\label{sec:conclusions}
%Grey-box recursive system identification is critical for accurately estimating key parameters in mineral flotation processes in real time. 
Our study evaluated the effectiveness of a recursive prediction error method in improving the accuracy of key performance indicators (KPIs) in a flotation model that incorporates froth physics. By tracking the parameter fluctuations due to operating conditions or disturbances, the proposed approach offers substantial benefits in terms of concentrate grade prediction compared to nominal parameter value predictions. The demonstrated reliability and precision of the recursive grey-box identification method in tracking and predicting concentrate grade suggests its broad applicability to other dynamic systems. Future work will focus on implementing a closed-loop model predictive control framework with the grey-box online identification technique to further optimize the froth flotation process.

%The mineral flotation model considered in this work
%Recursive estimation surpassed constant parameter models in all cases, indicating its superiority in dealing with real-world process variations. This approach offers substantial benefits for dynamic systems and is critical for informed decision-making and optimization in process control. 

%This approach offers substantial benefits for dynamic systems and is critical for informed decision-making and optimization in process control. 

%Our study evaluated the effectiveness of recursive estimation techniques in improving the accuracy of key performance indicators (KPIs) in a flotation system. We focused on three different scenarios: simultaneous changes in parameters $n$ and $C$, isolated changes in parameter $a$, and combined changes in parameters $b$ and $c$. Recursive estimation surpassed constant parameter models in all cases, indicating its superiority in dealing with real-world process variations. This approach offers substantial benefits for dynamic systems and is critical for informed decision-making and optimization in process control. 

%Future work will focus on implementing a closed-loop model predictive control framework with the grey-box online identification technique to further optimize the froth flotation process.
\bibliography{References}

%\newpage
%\appendix

\end{document}